\documentclass[%
 reprint,
 amsmath,amssymb,
 aps,
 prl,
]{revtex4-2}

\usepackage{graphicx}% Include figure files
\usepackage{dcolumn}% Align table columns on decimal point
\usepackage{bm}% bold math
\usepackage{xcolor}
\usepackage{url}

\begin{document}

%\preprint{APS/123-QED}

\title{Homophily impacts the success of vaccine roll-outs}% Force line breaks with \\

\author{Giulio Burgio}

\author{Benjamin Steinegger}%

\author{Alex Arenas}%

\email{alexandre.arenas@urv.cat}
\affiliation{
Departament d’Enginyeria Informàtica i Matemàtiques, Universitat Rovira i Virgili, 43007 Tarragona, Spain
}

\date{\today}% It is always \today, today,
             %  but any date may be explicitly specified

\begin{abstract}
Physical contacts do not occur randomly, rather, individuals with similar socio-demographic and behavioural characteristics are more likely to interact among them, a phenomenon known as homophily. Concurrently, the same characteristics correlate with the adoption of prophylactic tools. As a result, the latter do not unfold homogeneously in a population, affecting their ability to control the spread of infectious diseases. Here, focusing on the case of vaccines, we reveal three different dynamical regimes as a function of the mixing rate between vaccinated and non vaccinated individuals. Specifically, depending on the epidemic pressure, vaccine coverage and efficacy, we find the attack rate to decrease, increase or vary non monotonously with respect to the mixing rate. We corroborate the phenomenology through Monte Carlo simulations on a temporal physical contact network. Besides vaccines, our findings hold for a wide range of prophylactic tools, indicating a universal mechanism in spreading dynamics.

%We study how homophily of human interactions affect the impact of vaccines on the disease dynamics. Our mean-field analysis reveals three different dynamical regimes with respect to the mixing rate between vaccinated individuals and the ones that are not. We find that, depending on the epidemic pressure, vaccination level and efficacy, the attack rate can decrease, increase or vary non monotonously with respect to the mixing rate. We corroborate the phenomenology with Monte Carlo simulations on a temporal physical contact network. Further, besides vaccines, our findings hold for a wide range of prophylactic tools indicating a universal phenomena in spreading dynamics.      
\end{abstract}

%\keywords{Suggested keywords}%Use showkeys class option if keyword
                              %display desired
\maketitle

%\tableofcontents

%Necessity of vaccination
    %-best tools of humanity to combat viruses
    %-some almost eradicated through mandatory
%Not homogeneous 
    %-distribution not equal (worldwide, inside, opinion)
    %-voluntary vaccination implies clustering
    %-rising vaccines but clustering measles
    %-previous work analyzed this effect but for perfect vaccines as measles
    %-other examples such as influenza or Covid-19 (variants) imperfect

%Describe what we do
Vaccines have been crucial in humanity's struggle to protect itself from infectious diseases \cite{plagues}. In the 20th century, vaccines enabled to control a series of diseases \cite{Pollard2021} as well as the eradication of smallpox \cite{Smith2002}. Nevertheless, vaccines are not uniformly adopted in the population. While on a world-wide scale a lack of access impedes equitable adoption of vaccines \cite{Smith2011, Wouters2021}, among high income countries vaccine hesitancy is the primary barrier \cite{May2003, Parker2006, Health_2019}.      

Vaccine hesitancy widely correlates with age, socio-economic status, education level or ethnicity \cite{Robertson2021, Larson2014, MacDonald2015, Cascini2021}. Concurrently, these same factors shape the interaction patterns in society, leading to what is commonly referred to as homophily \cite{McPherson2001}, i.e., similarity of social contacts. As a consequence, social interactions are relatively homogeneous with regard to many socio-demographic or behavioral characteristics \cite{McPherson2001}, and vaccination status is not an exception \cite{May2003, Parker2006, Richard2008, Omer2008, Atwell2013, Barclay2014, Lieu2015, Edge2019}. This non-uniform, clustered vaccine adoption strongly determines how, and whether, the virus spreads in the population. A great example of this effect is provided by the recurrent measles outbreaks in high-income countries caused by clusters of vaccine hesitant individuals \cite{May2003, Parker2006, Richard2008, Omer2008, Atwell2013, Lieu2015}.  

%For example, the United States registered a series of outbreaks in 2019 \cite{patel2019national}, albeit the vaccine coverage being above the herd immunity level predicted assuming random mixing between vaccinated and non vaccinated individuals \cite{anderson1992infectious}. 

The recurring measles outbreaks sparked modelling studies that analyzed the impact of homophilic vaccine adoption on the disease dynamics \cite{Mbah2012, Salathe2008, Liu2015, Glasser2016, Kuylen2020}. Due to the high quality of vaccines against measles, these models assumed very high vaccine efficacy of almost 100\%, and showed that clustered adoption always increases the overall attack rate. In contrast, other vaccines such as the one against influenza or variants of concern of SARS-CoV-2 have relatively low efficacy, between 20-80\% \cite{CDCInfluenza, Pouwels2021}. 

In this work, through the joint exploration of imperfect immunization and vaccine uptake, we offer a wider picture in which vaccination clustering is not always detrimental. Specifically, we adapt a mean-field model recently proposed to study the impact of homophilic adoption of digital proximity tracing apps \cite{burgio2021homophily}, to analyse how the course of an epidemic is affected by the assortativity in the vaccination uptake. We show that the rich phenomenology found for tracing apps partially extends to the vaccination problem. Additionally, the findings also apply for a wide range of other prophylactic tools, eventually pointing out to a universal mechanism in spreading dynamics.
  
We consider the standard susceptible-infected-recovered (SIR) model, with transmission probability $\beta$, recovery rate $\mu$ and contact rate $k$. Accordingly, in the absence of protected individuals, and assuming homogeneous mixing, the basic reproduction number of the disease is $R_0 = \beta k/\mu$ \cite{anderson1992infectious}. The fraction of people who received a vaccine is fixed as $V\in\left[0,1\right]$. Upon encounter, an infected individual transmits the infection to a vaccinee at a reduced probability $\beta\left(1-\varepsilon\right)$, being $\varepsilon\in\left[0,1\right]$ the vaccine efficacy.  

Under random mixing, the effective reproduction number would be given by $R = R_0\left(1 - \varepsilon V\right)$. In order to include homophilic interactions, we now parametrize the mixing relation between vaccinated and non vaccinated people, i.e., the contact matrix $\mathbf{K}$, with a parameter $\alpha \in [0,1]$. We denote the entries of $\mathbf{K}$ as $k_{ij}$ with  $i,j \in \{\text{V}, \text{N}\}$, where $V$ and $N$ stand for \lq vaccinated' and \lq non vaccinated', respectively. We introduce $\alpha$ through the relation $k_{\text{NV}} = \alpha V k$, hence interpolating from complete homophily ($\alpha = 0$) to random mixing ($\alpha = 1$). The remaining contact rates follow from the balance equation $(1 - V) k_{\text{NV}} = V k_{\text{VN}}$ and the constraint $k = k_{\text{NN}} + k_{\text{NV}} = k_{\text{VV}} + k_{\text{VN}}$. Accordingly, $\mathbf{K}$ has the following entries 
\begin{align}
    k_{\text{NV}} &= \alpha V k \\
    k_{\text{NN}} &= \left[1 - \alpha V\right] k  \\
    k_{\text{VN}} &= \alpha (1 - V) k  \\
    k_{\text{VV}} &= \left[ 1 - \alpha (1 - V)\right]k  \,.
\end{align}

\noindent The degree of homophily regarding vaccine uptake, $h$, i.e., the probability that during a contact both individuals are either vaccinated or not, thus reads
\begin{equation}
    \label{eq:homophily}
    h = \frac{1}{k}\left[(1 - V) k_{\text{NN}} + V k_{\text{VV}}\right] = 1 - 2 \alpha V \left(1 - V\right)  \,.
\end{equation}

\noindent Note that $\alpha$ may take values larger than one (with an upper bound depending on $V$), indicating disassortative mixing. Broad sociological evidence \cite{McPherson2001,Centola2010spread,Centola2011}, however, largely excludes this possibility. Recently, the adoption of contact tracing apps has been shown to strongly correlate with age, income and nationality \cite{salathe2020early,Munzert2021,MorenoLopez2021}, effectively leading to homophilic (assortative) patterns. 

\begin{figure}[t]
    \includegraphics[width = 0.97\linewidth]{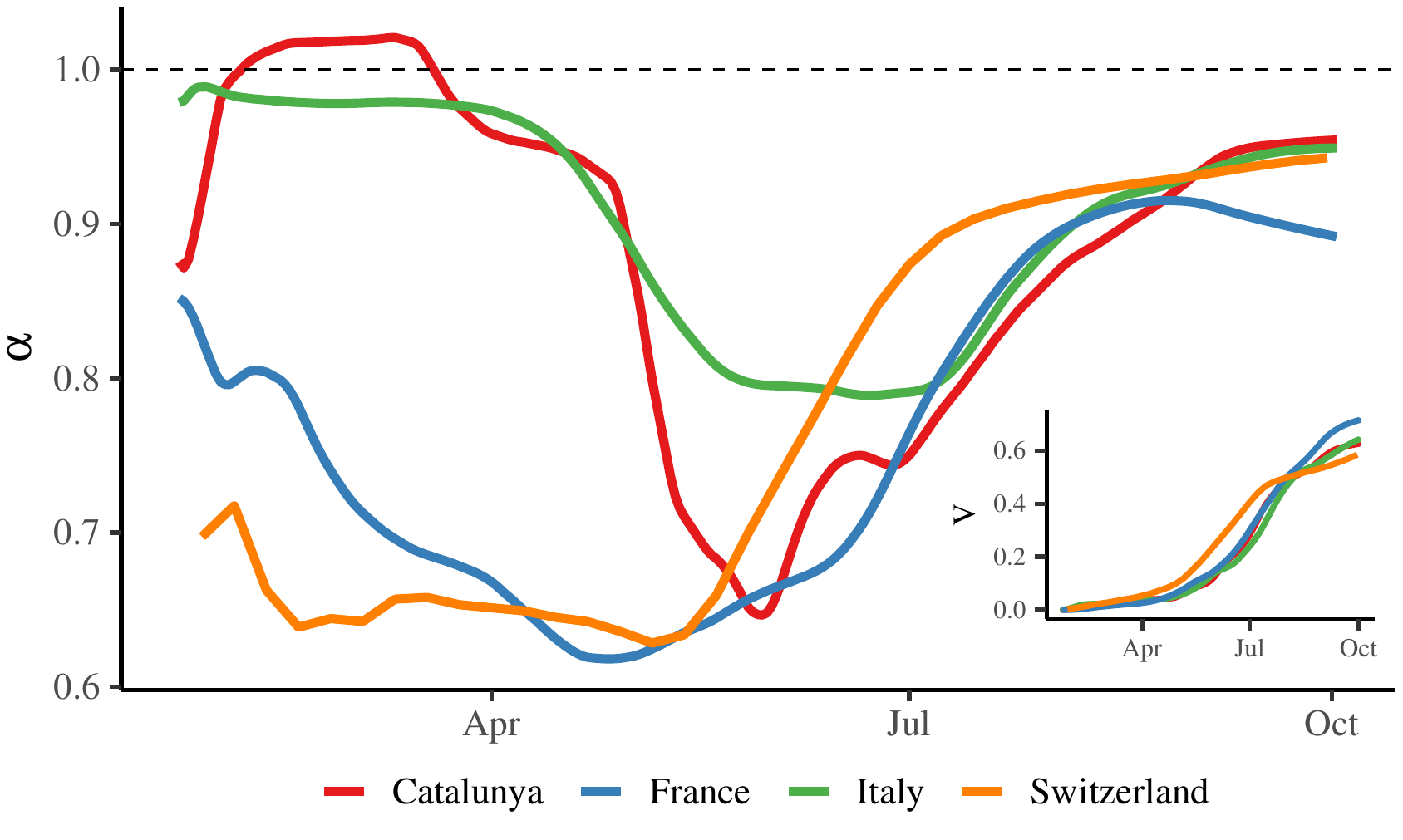}
    \caption{Weekly-moving average of the mixing parameter, $\alpha$, as inferred for different countries/regions starting from the contact matrix among age groups and the COVID-19 vaccines' uptake of each group (whose aggregate, $V$, is shown in the inset plot) from January to September 2021.}
    \label{fig:hom_from_ageStruct}
\end{figure}

Similarly, we tried to estimate assortativity in vaccine uptake indirectly from age-separated information. We leverage the correlations in the contact patterns among age groups and the levels of vaccine coverage within each age group. Specifically, for each region/country considered, we combined the age-separated (from $0$--$9$ to $70$--$79$, in steps of $10$ years) contact matrices \cite{prem2017projecting} with the data on the uptake of vaccines against COVID-19 (counting full vaccinations only), from January 2021 --when first full vaccinations appeared-- to September 2021 \cite{dataCAT,dataFR,dataIT,dataSWI}. As shown in Figure~\ref{fig:hom_from_ageStruct}, the estimated mixing rate stays well below $1$ for almost all the time, sometimes fluctuating around it only for very low levels of vaccine coverage ($V\lesssim 0.02$ for Catalonia). 

It must be said, however, that it would be naïve to take the temporal trends in Figure~\ref{fig:hom_from_ageStruct} as ready-to-use data. Indeed, we neglect other important features beyond age such as socio-economic classes or spatial patterns \cite{Robertson2021, Cascini2021}. As a matter of fact, we assume random mixing between vaccinated individuals and the ones that are not inside age groups. The results deriving from our simple analysis must be solely understood as a qualitative indication in line with existing literature reporting homophily in health behavior \cite{Barclay2014,Centola2011,MorenoLopez2021}, hence with the choice $\alpha\in[0,1]$. The latter is always kept fixed here, leaving the study of its implicit time-dependence for future work.

We denote with $X_{\text{Y}} \equiv X_{\text{Y}}(t)$ the fraction of vaccinated ($\text{Y = V}$) and not vaccinated ($\text{Y = N}$) people in compartment $X\in\left\{S,I\right\}$ at time $t$. Then, the differential equations governing the dynamics read as
\small
\begin{align}
    \dot{I_{\text{N}}}(t) &= \beta \left[ k_{\text{NN}}  I_{\text{N}}(t) + k_{\text{NV}} I_{\text{V}}(t)\right] S_{\text{N}}(t) - \mu I_{\text{N}}(t)  \label{eq:dyn6} \\ 
    \dot{I_{\text{V}}}(t) &=  \beta \left[ k_{\text{VN}}  I_{\text{N}}(t) + k_{\text{VV}} I_{\text{V}}(t)\right] (1 - \varepsilon) S_{\text{V}}(t) - \mu I_{\text{V}}(t) \\
    \dot{S_{\text{N}}}(t) &= -\beta \left[ k_{\text{NN}}  I_{\text{N}}(t) + k_{\text{NV}} I_{\text{V}}(t)\right] S_{\text{N}}(t) \\
    \dot{S_{\text{V}}}(t) &= -\beta \left[ k_{\text{VN}}  I_{\text{N}}(t) + k_{\text{VV}} I_{\text{V}}(t)\right] (1 - \varepsilon) S_{\text{V}}(t) \,.
    \label{eq:dyn9}
\end{align}
\normalsize

\begin{figure*}[t]
    \includegraphics[width = 1.0\linewidth]{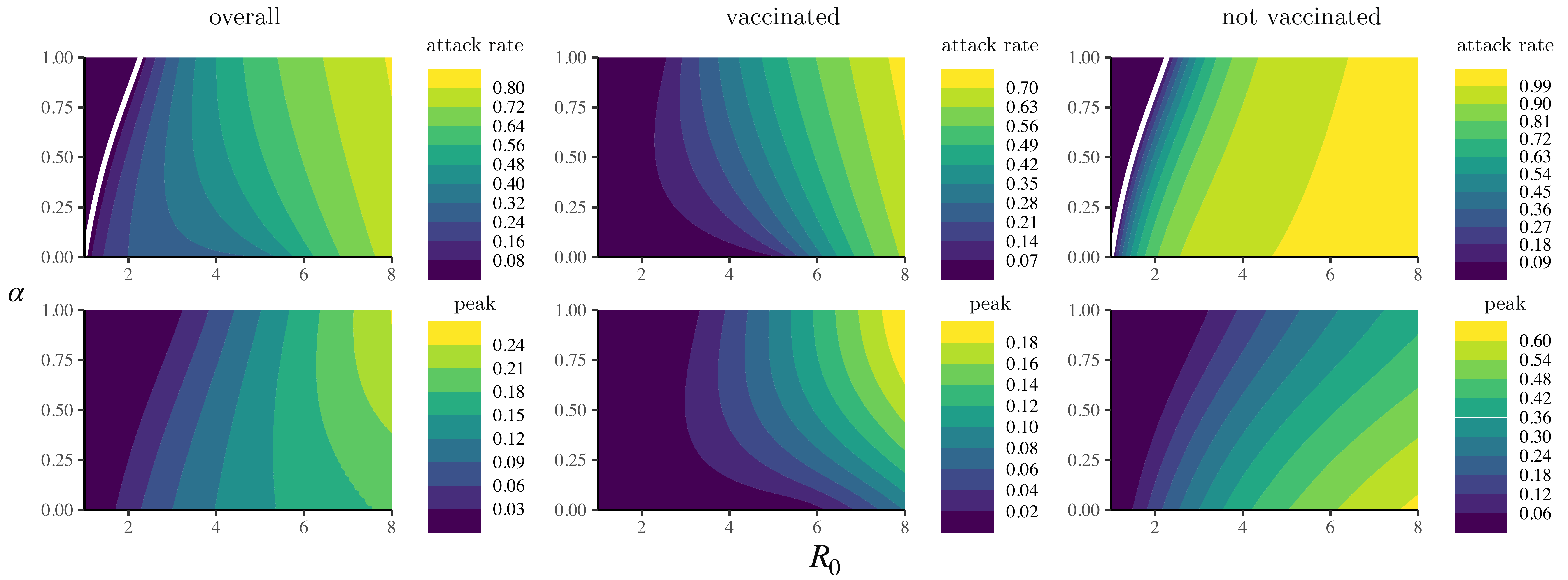}
    \caption{Attack rate (top) and peak of prevalence (bottom) as functions of the basic reproduction number, $R_0$, and the mixing parameter, $\alpha$, given $V = 0.7$ and $\varepsilon = 0.8$. In order to highlight the competing processes at the base of the dynamics we show, besides the results for the population overall, those for the vaccinated individuals and for the not vaccinated ones, separately. The solid line indicates the critical curve, $\alpha_c\equiv\alpha_c(R_0)$, at which $R=1$, as computed from Eq.~(\ref{eq:alpha_cr}).}
    \label{fig:heatmaps}
\end{figure*}

Linearizing around the disease-free equilibrium $(I_{\text{N}}, I_{\text{V}},S_{\text{N}}, S_{\text{V}}) \approx (0,0,1,1)$, the Jacobian matrix $\mathbf{J}$ of the system of Eqs.~(\ref{eq:dyn6})-(\ref{eq:dyn9}), takes the form
\begin{equation}
    \mathbf{J} = \begin{pmatrix}
    \beta k_{\text{NN}} - \mu & \beta k_{\text{NV}} \\
    \beta (1 - \varepsilon) k_{\text{VN}} & \beta (1 - \varepsilon) k_{\text{VV}} - \mu
    \end{pmatrix} \,.
\end{equation}

\noindent The effective reproduction number, $R$, is then found from the spectral radius $\rho(\mathbf{J})$ as $R = \rho(\mathbf{J}) + 1$ \cite{Diekmann1990}. If $\rho(\mathbf{J}) > 0$ ($R > 1$) the disease-free equilibrium is unstable and a finite fraction of the population gets infected. By computing the spectral radius and inserting the explicit expressions of the $\mathbf{K}$ matrix entries, one gets
\begin{align}
    \notag R =& \frac{R_0}{2}\left[2 - \alpha + \alpha\varepsilon (1-V) - \varepsilon ~+ \right. \\
    &\left.\sqrt{\left[\alpha-\alpha\varepsilon (1-V) + \varepsilon\right]^2 - 4\alpha \varepsilon V}\right] \,.
    \label{eq:repNumber}
\end{align}

\noindent One can easily check that $R/R_0$ decreases monotonously with respect to all the figuring parameters. The monotonous dependence on $\alpha$ was expected, as the higher is the latter, the lower is the number of effectively susceptible people an infected individual can come in contact with. This distinguishes the adoption of vaccines from contact tracing apps, for which $R$ can show instead a non monotonous dependence on the mixing rate \cite{burgio2021homophily}.

As anticipated, Eq.~(\ref{eq:repNumber}) reduces to $R = R_0\left(1 - \varepsilon V\right)$ for $\alpha = 1$, i.e., random mixing. Interestingly, the symmetry between efficacy, $\varepsilon$, and vaccine coverage, $V$, holding for random mixing, breaks for homophilic adoption (exception made for the degenerate case of complete homophily, i.e., $\alpha = 0$). Straightforward calculations prove that, kept fix the product $\varepsilon V$, a higher coverage lowers the reproduction number more than a higher efficacy (e.g., the pair of parameters $V=0.8$, $\varepsilon=0.6$ yields a lower $R$ than $V=0.6$, $\varepsilon=0.8$).

By imposing $R = 1$ and solving for $\alpha$, we find the critical value of mixing, $\alpha_c$, above which the disease cannot thrive, to be 
\begin{equation}
    \alpha_c = \left(1 - \frac{1}{R_0}\right)\frac{1-R_0(1-\varepsilon)}{1-R_0(1-\varepsilon) - \varepsilon (1-V)} \,,
    \label{eq:alpha_cr}
\end{equation}

\noindent given that $\alpha_c \geq  \left[2(1-1/R_0)-\varepsilon\right]/\left[1-\varepsilon(1-V)\right]$, with $R_0 > 1$, is satisfied. In particular, the condition is always met for $\varepsilon = 1$ and never met for $\varepsilon = 0$ (meaning eradication is not possible in this case, i.e., $R > 1$). The critical values of vaccine uptake, $V_c$, and efficacy, $\varepsilon_c$, are found from Eq.~(\ref{eq:alpha_cr}) by solving for the respective variable.

The attack rate, i.e., the final fraction of individuals that got infected (and eventually transitioned to the recovered/removed compartment), exhibits three different dynamical regimes with respect to its dependence on the mixing rate, $\alpha$. This is shown in the left panel of Figure~\ref{fig:heatmaps}, where, by increasing the basic reproduction number, $R_0$, the dependence of the attack rate on $\alpha$ is first monotonously decreasing, then concave and finally monotonously increasing. Following previous work \cite{burgio2021homophily}, we refer to these three regimes as critical, intermediate and saturated, respectively.

These three regimes result from the competition between two complementary, monotonous regimes, as illustrated in the central and right panels of Figure~\ref{fig:heatmaps}. With no mixing at all ($\alpha = 0$), vaccinated and not vaccinated form two disconnected components. Accordingly, the disease is free to spread in the not vaccinated cluster whenever $R=R_0>1$; on the other hand, $R=R_0(1-\varepsilon)<R_0$ in the vaccinated one, so the spread does not occur if $\varepsilon$ is high enough to ensure $R<1$. Increasing the mixing (i.e., $\alpha$), each cluster is `diluted' with nodes from the other. As a consequence, the reciprocal protection that vaccinated people provide among them to keep infection chains localized is partially lost. Instead, not vaccinated individuals profit from vaccinated individuals in their vicinity and are thus subjected to a lower infection probability. Mixing is therefore beneficial for not vaccinated people and detrimental for vaccinated ones. Then, depending on the basic reproduction number, $R_0$, the vaccine coverage, $V$, and the vaccine efficacy, $\varepsilon$, the overall system falls in one of the three dynamical regimes. Worthy to note, while the overall regimes can be explained in terms of that competition, the separate dynamics of vaccinated and not vaccinated compartments point to an even more complex quantitative behavior, as indicated by the change of slope of the contour curves in the central and right panels of Figure~\ref{fig:heatmaps}.

\begin{figure}[b]
    \includegraphics[width = 1.0\linewidth]{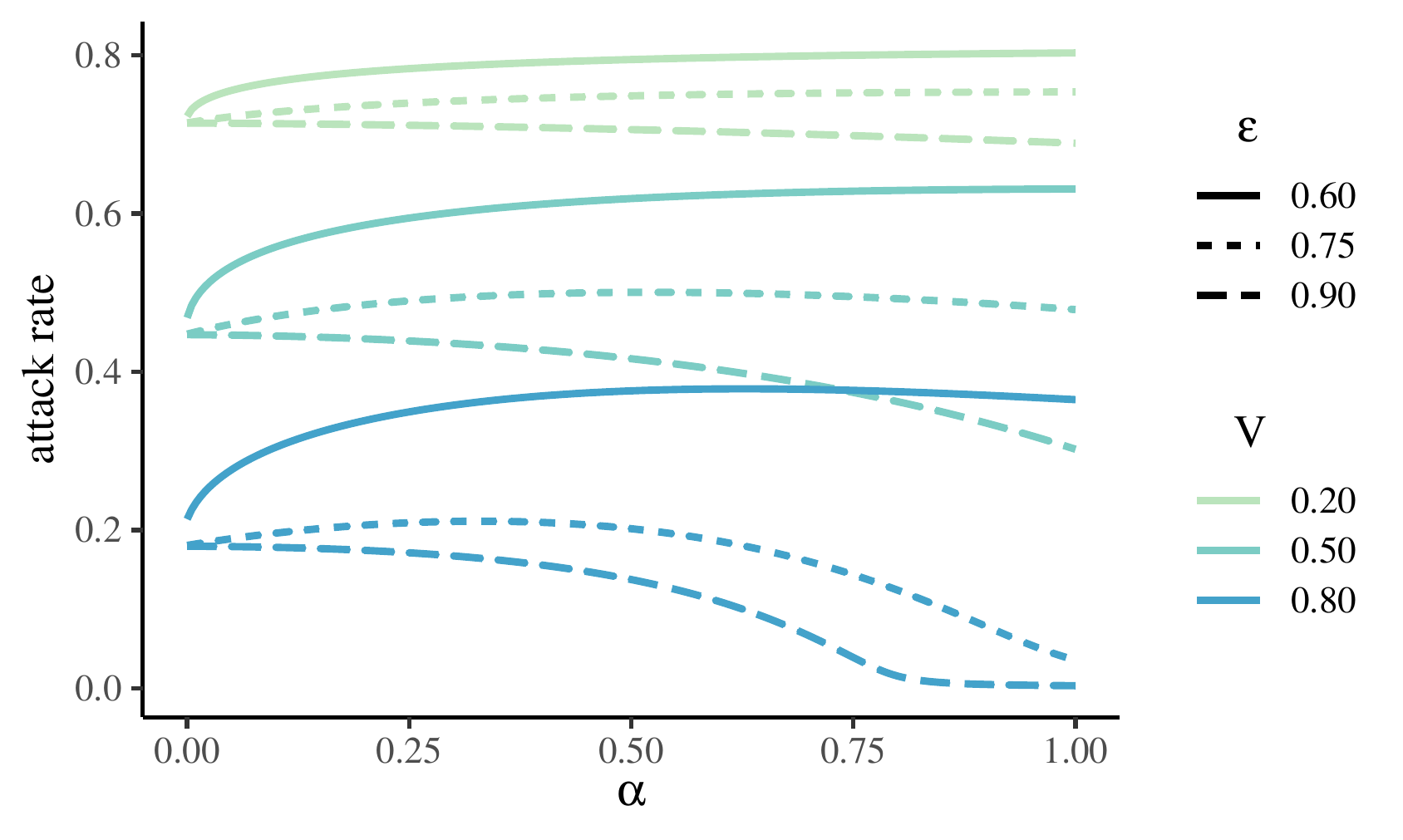}
    \caption{Attack rate as function of the mixing parameter, $\alpha$, for different combinations of vaccine coverage, $V$, and efficacy, $\varepsilon$, given $R_0 = 2.5$. As a complement to Figure~\ref{fig:heatmaps}, this plot illustrates how all three dynamical regimes can be also explored by varying $V$ or $\varepsilon$ (or both).}
    \label{fig:AR_vs_mixing}
\end{figure}

The same qualitative picture holds for the peak in prevalence (see bottom panels of Figure~\ref{fig:heatmaps}), i.e., the maximum number of simultaneously infectious people during the epidemic. Note that, for each set of fixed parameters, the respective regimes of the peak in prevalence and of the attack rate may not coincide. For example, at $R_0=5$, the attack rate is in the saturated regime while the peak in prevalence is still in the critical regime (Figure~\ref{fig:heatmaps}). In Figure~\ref{fig:AR_vs_mixing}, we show how the three regimes can be also accessed by varying the vaccine coverage, $V$, or the vaccine efficacy, $\varepsilon$, or both. Specifically, increasing $V$ and/or $\varepsilon$ makes the dynamics pass from the saturated to the critical regime, going through the intermediate one. Accordingly, a sufficiently high efficacy makes random mixing always beneficial.  

In the case of a perfect vaccine, i.e., $\varepsilon = 1$, the system is always in the critical regime. Indeed, in this case, we see from Eqs.~(\ref{eq:dyn6})-(\ref{eq:dyn9}) that the dynamics reduces to a standard SIR model within the non vaccinated sub-population ($S_{\text{N}}$, $I_{\text{N}}$). Such dynamics is then solely driven by the contact rate $k_{\text{NN}}$, which is a linearly decreasing function of the mixing rate, $\alpha$.

At this point, we are ready to link the existence of the three dynamical regimes revealed for both vaccination and contact tracing apps \cite{burgio2021homophily} to a common basic mechanism. The essential ingredient of the saturated regime is the lack of sufficient individual, independent protection offered by the prophylactic measure to the adopter. This is evident for contact tracing apps, as adoption does not directly protect the adopter. Instead, individuals only indirectly benefit if adoption is sufficiently widespread such that transmission chains between adopters can be stopped. Similarly, if the vaccine is imperfect, isolated adoption in a high prevalence environment may not substantially reduce the individual infection probability. Instead, widespread adoption among an individual's contacts is necessary to provide mutual protection.   

Accordingly, for both vaccines and contact tracing apps, if coverage is low in comparison to the epidemic pressure, only an assortative/clustered adoption can provide mutual protection and thus non adopters cannot be protected (saturated regime). On the contrary, if coverage is high, protection is sufficiently strong such that non adopters can be protected and thus mixing between adopters and non adopters is beneficial (critical regime). In between we then find the intermediate regime. In all three cases, this phenomenology holds for any prophylactic measure that reduces the transmission probability since it is mathematically equivalent to the model presented here. In this sense, equivalent results could be found for the use of face masks or the adoption of social distancing. 

\begin{figure}[t]
    \includegraphics[width = 1.0\linewidth]{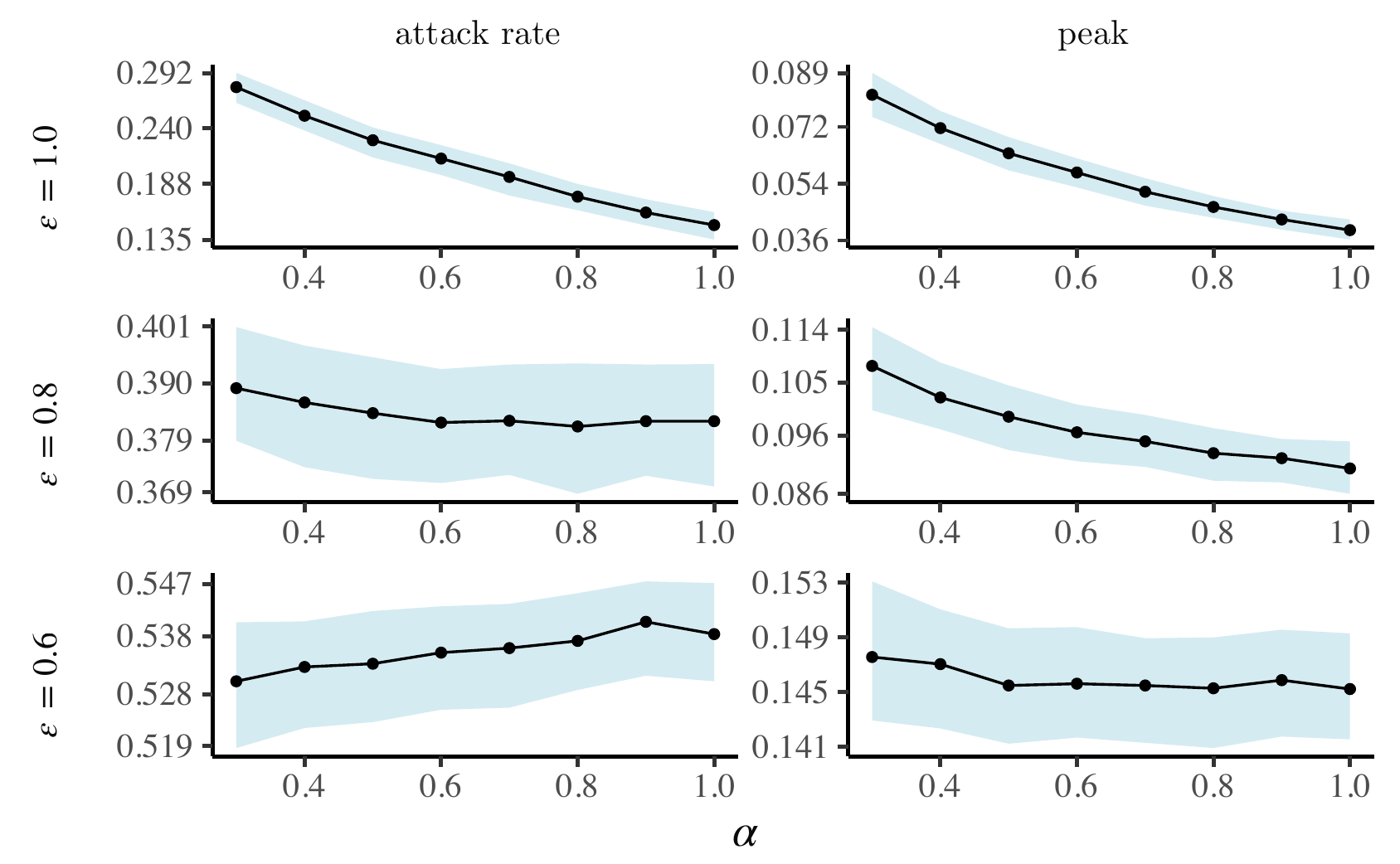}
    \caption{Attack rate (left) and peak of prevalence (right) as functions of the mixing parameter, $\alpha$, for decreasing (from top to bottom) efficacy, $\varepsilon$, as resulting from the numerical simulations performed on top of a real-world temporal contact network (see the main text for details). Dots indicate the median value, whereas the ribbon indicates first and third quartiles. Each point is obtained by averaging over $2\times10^4$ runs. We fixed $V = 0.5$, $\mu = 4.6\times10^{-4}$ (corresponding to a mean infection time of $7.5$ days) and $\beta = 1.142\times10^{-1}$, yielding $R_0=6$ from the estimation $R_0 = \beta\kappa/\mu$, where $\kappa=\overline{s^2}/\bar{s}$, being $\bar{s}$ ($\overline{s^2}$) the network average of the (squared) number of contacts per timestamp.}
    \label{fig:res_TCN}
\end{figure}

To corroborate our theory, we performed numerical simulations upon a temporal contact network estimated via Bluetooth signal exchanges in the Copenhagen Networks Study \cite{Sapiezynski_2019}. We retained those interactions with an associated Received Signal Strength Indication (RSSI) not lower than $-74$ dBm, corresponding to physical distances approximately up to $2$ meters \cite{sekara2014strength}. The resulting temporal network involves $672$ individuals and $374884$ pairwise interactions spread over $8064$ timestamps, binning four weeks of recording time into five-minute intervals. 

To distribute the vaccine in the population, we provisionally aggregate the temporal network to get a weighted, static one, where the weight of an edge is the number of times its end nodes interacted. Accordingly, the homophily, $h$, is computed as the sum of the weights (duration of contacts) over the homophilic edges, normalized by the sum over all the edges. We initially distribute the vaccine at random ($\alpha\approx 1$) and then iteratively swap the vaccination status of two randomly selected neighboring nodes whenever this leads to a decrease in $\alpha$, until the latter attains a preset value ($\pm 0.01$). Additionally, we control that both adopter and non adopters have the same average degree. Otherwise, the algorithm induces a spurious correlation between vaccination status and number of contacts of a node. Given the role that highly connected nodes play in driving the spreading dynamics, such correlation can importantly affect the results. Therefore, as done in \cite{burgio2021homophily} --but not in a previous, related work \cite{Barclay2014}--, we allow only innocuous, not correlating swaps to be performed. 

The results, reported in Figure~\ref{fig:res_TCN}, confirm the existence of the three dynamical regimes identified by our model. Accordingly, the beneficial effect of random mixing vanishes when the vaccine does not provide sufficient protection. In the saturated regime, mixing is slightly detrimental. Furthermore, as in the mean-field case, we observe that final attack rate and peak of prevalence can be in different dynamical regimes (see bottom panels of Figure~\ref{fig:res_TCN}). Particular structural constraints of this aggregated network do not allow for arbitrarily small values of $\alpha$, making the saturated regime less evident. Nonetheless, higher homophilic levels (even $\alpha \approx 0$) can be reached by artificially controlling the mixing, in which case the decrease in the attack rate would be better appreciated.

Depending on the need and the available information, the model can be applied to different specific populations and at different scales. Currently, many epidemic models only implictily consider the mixing between not vaccinated and vaccinated individuals through stratification according to age (Fig.~\ref{fig:hom_from_ageStruct}) \cite{Bubar2021, Sonabend2021} or socio-economic status \cite{Brand2021}. Our findings urge also to account for the impact that subgroup-specific levels of mixing, can have on the system as a whole. Such approaches may provide a further tool to interpret epidemiological data. Additionally, more realistic models can act as useful guides for policy makers in cases where the mixing is (partially) controllable or can be influenced, at least. For instance, with respect to the current COVID-19 pandemic, the requirement in various countries of the green pass to enter restaurants or the workplace strongly reshapes the social fabric \cite{Dada2021}. In light of our results, the associated reduction in the mixing level between not vaccinated individuals and vaccinated ones may substantially alter the epidemic dynamics. Obviously, the aim of vaccines is not only to prevent infections but as well to reduce the number of severe cases \cite{Bubar2021}. Additionally, the intention behind the green pass is to nudge individuals to get vaccinated or to reduce their number of physical contacts \cite{Wilf2021}. Nevertheless, our results indicate that more detailed information on the correlations between social interactions and health behaviour (vaccination, face mask, social distancing, contact tracing apps) would lead to a more comprehensive analysis. 

From a more theoretical standpoint, we showed that the presence of homophily in vaccine adoption leads to three different dynamical regimes (critical, intermediate, saturated). Furthermore, the phenomenology presented here also extend to any prophylactic measure that reduces the transmission probability (face masks, social distancing) as well as digital contact tracing \cite{burgio2021homophily, rizi2021epidemic}. Accordingly, the phenomenology induced by the presence of homophily is robust with respect to the adoption of different health behaviours. This robustness of our findings across prophylactic tools hints to a general feature of spreading dynamics. Eventually, the results highlight how correlated metadata such as vaccination status can add rich phenomenology beyond the network structure itself \cite{Gomez_2011, Newman2016, Peel2017, Artime2021}. \\

%In this sense, we assume a leaky instead of an "all-or-nothing" vaccine \cite{Halloran2010}. While the former reduces the infection probability on a per-exposure basis, the latter provides complete protection to a subset of the vaccinated individuals independent of exposure. Later in the manuscript, we will comment on how the results change when considering an "all-or-nothing" vaccine.
%given that risk is much higher for elder people, reducing their mixing with younger and less vaccinated age groups can be a helpful strategy, even when this, if coupled to a low vaccine efficacy, may lead to an increase in infections in the overall population. Meanwhile, within a young age group, separating vaccinated from not vaccinated people may be useless (or even deleterious) when vaccine efficacy is high, as the age group as a whole could not benefit from the herd protection offered by the vaccinee; yet may be convenient when efficacy is low. The same can be said for those group activities requiring many individuals working physically together: depending on the vaccine efficacy (and assuming workers are somewhat interchangeable), may be safer to cluster workers depending on their vaccination status or leave them mixed as in an ordinary situation.

\emph{Note added.} Few days after having finished writing this article, we became aware that Hiraoka et al. made an analysis very similar to ours \cite{hiraoka2021herd}. The phenomenology they uncover for random networks is equivalent to the one we found in the mean field case and on a real temporal physical contact network. This further illustrates the robustness of the different dynamical regimes. \\

\emph{Acknowledgments.} G.B. acknowledges financial support from the European Union's Horizon 2020 research and innovation program under the Marie Sk\l{}odowska-Curie Grant Agreement No. 945413 and from the Universitat Rovira i Virgili (URV). B.S. acknowledges financial support from the European Unions Horizon 2020 research and innovation program under the Marie Sk\l{}odowska-Curie Grant Agreement No. 713679 and from the Universitat Rovira i Virgili (URV). A.A. acknowledges support by Ministerio de Econom\'ia y Competitividad (Grants No. PGC2018-094754-B-C21 and No. FIS2015-71582-C2-1), Generalitat de Catalunya (Grant No. 2017SGR-896), Universitat Rovira i Virgili (Grant No. 2019PFR-URV-B2-41), ICREA Academia, and the James S. McDonnell Foundation (Grant No. 220020325). We thank L. Arola-Fern{\'a}ndez for helpful comments and suggestions.

\bibliography{apssamp}

\end{document}